\documentclass[prd,longbibliography,superscriptaddress,showpacs,showkeys,twocolumn,floatfix,eqsecnum]{revtex4-1}
\usepackage{textcomp}
\usepackage{eurosym}
\usepackage{amsfonts}
\usepackage{changes}
\usepackage{url}
\usepackage{array}
\usepackage{amsthm}
\usepackage{bm}
\usepackage{palatino}
\usepackage[colorinlistoftodos]{todonotes}
\usepackage{mathpazo}
\usepackage{supertabular}
\usepackage{subfig}
\usepackage{amssymb}
\usepackage{eurosym}
\usepackage{amsmath}
\usepackage{graphics}
\usepackage{color}
\usepackage{graphicx}
\usepackage[font={footnotesize,it}]{caption}
\usepackage[colorlinks=true,
            linkcolor=blue,
            urlcolor=blue,
            citecolor=green]{hyperref}
\usepackage{tabstackengine}
\usepackage{graphicx}
\usepackage[english]{babel}
\usepackage{amsfonts}
\usepackage{eurosym}
\usepackage{calrsfs}
\usepackage{color}

\def\be{\begin{equation}}
\def\ee{\end{equation}}

\begin{document}

\title{Exact traversable wormhole solution in bumblebee gravity}
\author{Ali \"{O}vg\"{u}n}
\email{ali.ovgun@pucv.cl}
\homepage{https://aovgun.weebly.com} 
\affiliation{Instituto de F\'{\i}sica,
Pontificia Universidad Cat\'olica de Valpara\'{\i}so, Casilla 4950,
Valpara\'{\i}so, Chile.}

\affiliation{Physics Department, Arts and Sciences Faculty, Eastern Mediterranean
University, Famagusta, North Cyprus via Mersin 10, Turkey.}

\author{Kimet Jusufi}
\email{kimet.jusufi@unite.edu.mk}

\affiliation{Physics Department, State University of Tetovo, Ilinden Street nn, 1200,
Tetovo, Macedonia.} 
\affiliation{Institute of Physics, Faculty of Natural Sciences and Mathematics, Ss. Cyril
and Methodius University, Arhimedova 3, 1000 Skopje, Macedonia.}

\author{\.{I}zzet Sakall{\i}}
\email{izzet.sakalli@emu.edu.tr}
\affiliation{Physics Department, Arts and Sciences Faculty, Eastern Mediterranean
University, Famagusta, North Cyprus via Mersin 10, Turkey.}

\begin{abstract}
In this study, we found a new traversable wormhole solution in the framework
of a bumblebee gravity model. With these types of models, the Lorentz
symmetry violation arises from the dynamics of a bumblebee vector field that
is non-minimally coupled with gravity. To this end, we checked the
wormhole's flare-out and energy (null, weak, and strong) conditions. We then
studied the deflection angle of light in the weak limit approximation using
the Gibbons-Werner method. In particular, we show that the bumblebee gravity
effect leads to a non-trivial global topology of the wormhole spacetime. By
using the Gauss-Bonnet theorem (GBT), it is shown that the obtained
non-asymptotically flat wormhole solution yields a topological term in the
deflection angle of light. This term is proportional to the coupling
constant, but independent from the impact factor parameter. Significantly,
we showed that the bumblebee wormhole solutions, under specific conditions,
support the normal matter wormhole geometries.
\end{abstract}

\date{\today}
\maketitle

\section{Introduction}

\begin{figure}[h!]
\center
\includegraphics[width=0.5\textwidth]{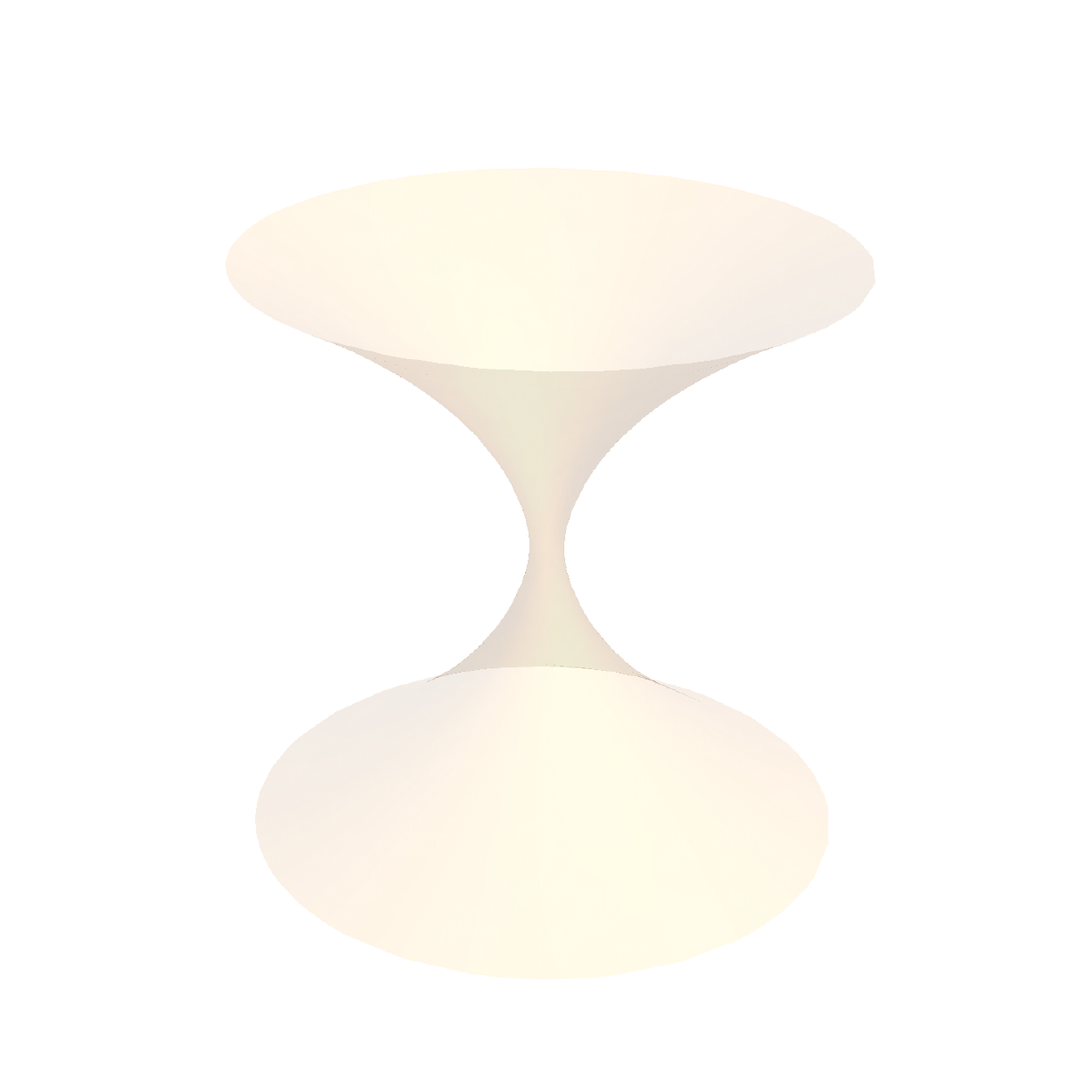}
\caption{Traversable Wormhole.}
\label{w1}
\end{figure}

The search for a theory of wormholes through Einstein's general theory of
relativity goes back to 1916 with the famous papers of Flamm \cite{Flamm}.
The simplest possible solution to Einstein's field equations is the
Schwarzschild metric \cite{Schwz}, which describes the gravitational field
around a spherically symmetric static mass. If the mass (or its density) is
sufficiently high, the solution describes a black hole -- the Schwarzschild
black hole. Flamm realized that Einstein's equations also allow a second
solution, which is presently known as a \textquotedblleft white
hole\textquotedblright. These two solutions, describing two different
regions of (flat) spacetime, are connected by a \textquotedblleft spacetime
tube\textquotedblright. This tube does not define where those regions of
spacetime might be in the real world; the black hole's \textquotedblleft
entrance\textquotedblright and white hole's \textquotedblleft
exit\textquotedblright could exist in different portions of the same
universe or in entirely different universes. In 1935, Einstein and Rosen 
\cite{ERb} further explored the theory of inter-universe connections. In
fact, their main aim was to try to understand the fundamental charged
particles (protons, electrons, etc.) in terms of spacetime tubes penetrated
by lines of electromagnetic force. These spacetime passageways were named
\textquotedblleft Einstein-Rosen Bridges\textquotedblright\ by Wheeler, who
would later call them \textit{wormholes}. It is worth noting that Wheeler 
\cite{Wheeler} also coined the term \textquotedblleft black
hole.\textquotedblright\ Traversable wormholes have no horizon and allow
two-way traveling \cite{Travel} by connecting two different regions of
spacetime in a Lorentzian geometry. Interest in traversable wormhole gained
momentum following the paper of Morris, Thorne, and Yurtsever (MTY) \cite%
{MTY} as shown in Fig. \ref{w1}. With a traversable wormhole, an
interstellar or inter-universe journey is possible \cite{Travel2,Visser}.
However, to construct such a traversable wormhole, one requires an exotic
matter with a negative energy density and a large negative pressure, which
should have a higher value than the energy density. Meanwhile, the Casimir
effect \cite{Casimir} is a way of producing negative energy density. MTY
also proved that traversable wormholes could be stabilized using the Casimir
effect. Toward this end, placing two sufficiently charged superconducting
spheres at the traversable wormhole mouths is enough. On the other hand, in
2011, Kanti and Kleihaus \cite{Kanti} showed that it might be possible to
construct a traversable wormhole using normal matter by resorting to a form
of string theory.

In the literature, many authors have intensively studied various aspects of
traversable wormhole geometries within different modified gravitational
theories \cite%
{Hochberg:1996ee,Harko:2013aya,Lemos:2003jb,Lobo:2005us,Lobo:2005us,Lobo:2009ip,Bohmer:2011si,Zangeneh:2015jda,Clement:1983fe,Clement:1994qb,Susskind:2017nto,Baez:2014bka,Maldacena:2013xja,Poisson:1995sv,Hochberg:1997wp, Bronnikov:2002rn,Bronnikov:2009na,Bronnikov:2010tt,Garattini:2007ff,Cataldo:2017ard,Cataldo:2016dxq,Cataldo:2015vra,Cataldo:2017yec,Ovgun:2018uin,Richarte:2017iit,Ovgun:2017jip,Halilsoy:2013iza,Sakalli:2015taa,Sakalli:2015mka,Ovgun:2015una,Ovgun:2015sqa,Ovgun:2016ijz,Ovgun:2016ujt,Kim:1992sh,Sushkov:2002ef,Kim:2001ri,Guendelman:2011qj,Guendelman:2010pr,Guendelman:2009zz,Guendelman:2009pf,Guendelman:1991pc,MontelongoGarcia:2010xd,Bhar:2016vdn,Rahaman:2016jds,Bhar:2014ooa,Rahaman:2014pba,Rahaman:2014dpa,Kuhfittig:2013hva,Rahaman:2013xoa,Jamil:2013tva,Rahaman:2012pg,Eling:2006df}%
. Among them, the bumblebee gravity model has dynamically violated Lorentz
symmetry in terms of charge conjugation, parity transformation, and time
reversal. The model, with its defined bumblebee vector field, can also
feature rotation and boost \cite%
{Bluhm:2004ep,Bertolami:2005bh,Bluhm:2008yt,Seifert:2009gi,Kostelecky:2010ze,Escobar:2017fdi}%
. In fact, bumblebee gravity was first used by Kostelecky and Samuel in 1989 
\cite{bumb1,namb} as a simple model for investigating the consequences of
spontaneous Lorentz violation. 

The bumblebee mechanism arose in the context of string theory and lead to a
spontaneous breaking of Lorentz symmetry by tensor-valued fields acquiring
vacuum expectation values \cite{Kostelecky:2010ze}. The forcefulness of the
bumblebee vector field on the gravitational field has motivated us to
construct traversable wormholes. Very recently, a Schwarzschild-like
bumblebee black hole solution has been obtained \cite{bumblebee}. From the
perspective of string theory and loop quantum gravity theory, Lorentz
symmetry breaking (LSB) is an interesting idea for exploring the tracks of
the quantum gravity at low energy levels. LSB has been extensively studied
in the literature, e.g., see \cite%
{LSB,Maluf:2015hda,Capelo:2015ipa,Nascimento:2014vva,Maluf:2014dpa,Paramos:2014mda}
and references therein.

The main aim of this paper is to construct an exact solution of a
traversable wormhole in the bumblebee gravity field, where in Einstein's
field equations are influenced by spontaneous breaking of Lorentz symmetry.
We sought to compute the weak deflection angle of the obtained bumblebee
wormhole. To this end, we employed the Gibbons-Werner method (GWM) \cite{gibbons}.
In this method, the deflection angle, for the weak lensing limit, is
calculated by the GBT, defined by the background optical
geometry. It is important to highlight that non-singular domains are
considered to be outside of the light ray, which means that the GBT has a global impact \cite{gibbons,werner,Bonnet,gibbons3}. Wormholes
have been widely studied by many authors as have black holes; light
deflection has been of particular interest \cite%
{kimet22,wh10,Jusufi:2017mav,kimet1,kaa,kimet3,Jusufi:2017drg,Jusufi:2017xnr,aovgun,Ovgun:2018tua,Ovgun:2018ran,Ovgun:2018fnk,asahi1,Ono:2017pie,Crisnejo:2018uyn,asada,blackholes-wormholes1,rajibul,rajibul2,rajibul3}%
. Another purpose of this paper is to discover a traversable wormhole using
normal matter, which satisfies energy and flare-out conditions in the
bumblebee gravity. In the following sections, we shall explain how these
goals are achieved.

This paper is organized as follows: In Sec. II, we briefly outline bumblebee
gravity and its corresponding Einstein's field equations. In Sec. III, we
present LSB wormhole solutions and study the flare-out conditions. We check
the energy conditions of the bumblebee wormhole in Sec. IV. In the framework
of the GWM, Sec. V is devoted to the study of the
deflection angle of light in the weak limit approximation. Our conclusions
and remarks follow in Sec. VI.

\section{Bumblebee gravity}

The action of the bumblebee gravity where the Lorentz violation arises from
the dynamics of a single vector field, namely bumblebee field $B_{\mu }$,
with a real coupling constant $\xi$ (with mass dimension --1) is given by

\begin{eqnarray}
S_{B} &=&\int d^{4}x\sqrt{-g}\,\Big[\frac{1}{2\,\kappa }R+\frac{1}{2\,\kappa 
}\xi \,{B^{\mu }\,B^{\nu }}R_{\mu \nu }  \notag \\
&-&\frac{1}{4}B_{\mu \nu }B^{\mu \nu }-V(B^{\mu })\Big]+\int d^{4}x\,%
\mathcal{L}_{\text{M}},  \label{actiown}
\end{eqnarray}%
where the bumblebee field strength ($B_{\mu \nu }$) and the potential ($V$)
are defined as follows 
\begin{equation}
B_{\mu \nu }=\partial _{\mu }B_{\nu }-\partial _{\mu }B_{\nu },  \label{bee1}
\end{equation}

\begin{equation}
V\equiv V(B^{\mu }B_{\mu }\pm a^{2}).  \label{Vw}
\end{equation}

where $a^{2}$ is a positive real constant
\cite{bumblebee}. Vacuum expectation value of the bumblebee gravitational
field is governed by the following condition

\begin{equation*}
V(B^{\mu }B_{\mu }\pm a^{2})=0.
\end{equation*}

This automatically implies that {\ }%
\begin{equation}
B^{\mu }B_{\mu }\pm a^{2}=0\text{,}  \label{con}
\end{equation}%
in which the signs ($\pm )$ potentially determines {the field
type of }$b_{\mu }:$ timelike or spacelike. Solutions of Eq. (\ref{con}) {%
are conditional on the field $B^{\mu }$ that acquires a non-null vacuum
expectation value: } 
\begin{equation}
\left\langle B^{\mu }\right\rangle =b^{\mu }\text{.}  \label{bB}
\end{equation}%
In this setup, we use null torsion and null cosmological constant, so that there is a non-null vector $b^{\mu }$ which satisfies $b^{\mu }b_{\mu }=\mp 
a^{2}=\alpha=\text{constant}$. Thus, the nonzero vector background $b^{\mu }$, which is a coefficient for Lorentz and CPT violation, spontaneously
breaks the $U(1)$ symmetry \cite{Alan2004}.

Bumblebee modified Einstein field equations \cite{bumblebee} are governed by 
\begin{equation}
G_{\mu \nu }=\kappa T_{\mu \nu },  \label{EEQ}
\end{equation}

where the total energy-momentum tensor is given by \cite{Alan2004}

\begin{equation}
T_{\mu \nu }=T_{\mu \nu }^{M}+T_{\mu \nu }^{B},
\end{equation}

in which $T_{\mu \nu }^{M}$ is the matter field and the bumblebee
energy-momentum tensor $T_{\mu \nu }^{B}$ reads

\begin{eqnarray}
T_{\mu \nu }^{B} &=&-B_{\mu \alpha }B_{\nu }^{\alpha }-\frac{1}{4}B_{\alpha
\beta }B^{\alpha \beta }g_{\mu \nu }-Vg_{\mu \nu }+2V^{\prime }B_{\mu
}B_{\nu }  \notag \\
&+&\frac{\xi }{\kappa }\Big[\frac{1}{2}B^{\alpha }B^{\beta }R_{\alpha \beta
}g_{\mu \nu }-B_{\mu }B^{\alpha }R_{\alpha \nu }-B_{\nu }B^{\alpha
}R_{\alpha \mu }  \notag \\
&+&\frac{1}{2}\nabla _{\alpha }\nabla _{\mu }(B^{\alpha }B_{\nu })-\frac{1}{2%
}\nabla ^{2}(B_{\mu }B_{\nu })  \notag \\
&-&\frac{1}{2}g_{\mu \nu }\nabla _{\alpha }\nabla _{\beta }(B^{\alpha
}B^{\beta })\Big].
\end{eqnarray}%
Thus, the modified Einstein field equations (\ref{EEQ}) with the bumblebee
field can be expressed as follows \cite{bumblebee} 
\begin{equation}
R_{\mu \nu }-8\pi G\left[ T_{\mu \nu }^{M}+T_{\mu \nu }^{B}-\frac{1}{2}%
g_{\mu \nu }\left( T^{M}+T^{B}\right) \right] =0,
\end{equation}%
which has the following explicit form: 
\begin{eqnarray}
E_{\mu \nu }^{instein} &=&R_{\mu \nu }-\kappa \left( T_{\mu \nu }^{M}-\frac{1%
}{2}g_{\mu \nu }T^{M}\right) -\kappa T_{\mu \nu }^{B}-2\kappa g_{\mu \nu }V 
\notag \\
&+&\kappa B_{\alpha }B^{\alpha }g_{\mu \nu }V^{\prime }-\frac{\xi }{4}g_{\mu
\nu }\nabla ^{2}(B_{\alpha }B^{\alpha })  \notag \\
&-&\frac{\xi }{2}g_{\mu \nu }\nabla _{\alpha }\nabla _{\beta }(B^{\alpha
}B^{\beta })=0,  \label{meins1}
\end{eqnarray}%
where $T^{M}=g^{\mu \nu }T_{\mu \nu }^{M}$ and $\kappa =8\pi $. %
The prime denotes a derivative with respect to $r$. It can
be easily checked that when the both bumblebee field $B_{\mu }$ and
potential $V(B_{\mu })$ are vanished, the original general relativity field
equations are recovered.

Here, we focus on the vacuum solutions induced by the LSB, which is possible
when the bumblebee field $B_{\mu }$ remains frozen in its vacuum expectation
value. Namely, we consider the case of Eq. (\ref{bB}), so that we have a
vanishing potential: $V=V^{\prime }=0$ \cite{vacbump}. Thus, the bumblebee
field strength (\ref{bee1}) becomes 
\begin{equation}
b_{\mu \nu }=\partial _{\mu }b_{\nu }-\partial _{\nu }b_{\mu }.
\end{equation}

\section{Exact solution of bumblebee wormhole}

In this section, we consider a static and a spherically symmetric
traversable wormhole solution \cite{Travel} without any tidal force 
\begin{equation}
ds^{2}=-dt^{2}+\frac{dr^{2}}{1-\frac{W(r)}{r}}+r^{2}d\theta ^{2}+r^{2}\sin
^{2}\theta d\varphi ^{2},  \label{mymetric}
\end{equation}%
where $W(r)$ is the shape function of the wormhole. Furthermore, we set the
bumblebee vector as follows 
\begin{equation}
b_{\mu }=\left[ 0,\sqrt{\frac{\alpha }{1-\frac{W(r)}{r}}},0,0\right]. \label{mymbumb}
\end{equation}

The bumblebee modified Einstein's field equations with the isotropic matter $%
({{T^{\mu }}_{\nu }})^{M}=(-\rho ,p,p,p)$ \cite{MontelongoGarcia:2010xd}. We
shall use the equation of state: $p=w\rho ,$ in which $\rho $ denotes the
energy density of the matter field, $p$ stands for the pressure, and $w$ is
the dimensionless $\mathbb{R}$ number. Thus, Eq. (\ref{meins1}) yields the
following three Einstein's field equations in the bumblebee gravity theory
for the wormhole metric (\ref{mymetric}) (the details are tabulated in Appendix):

\begin{widetext}
\begin{eqnarray} \label{3.4}
E_{t t }^{instein}&=& -\kappa\,\rho\,{r}^{3}-3\,\kappa\,w\,\rho\,{r}^{3}+\xi\,\alpha\,
rW^{\prime}-\xi\,\alpha\,W \left( r
 \right) =0
,\\ \label{3.5}
E_{r r }^{instein}&=& 2\, rW^{\prime}-2\,W
 \left( r \right) +\kappa\,w\,\rho\,{r}^{3}-\kappa\,\rho\,{r}^{3}
+3\,\xi\,\alpha\,rW^{\prime} -3\,
\xi\,\alpha\,W \left( r \right) =0
,\\ \label{3.6}
E_{\theta \theta }^{instein}&=& \kappa\,w\,\rho\,{r}^{3}-\kappa\,\rho\,{r}^{3}+2\,\xi\,\alpha\,W
 \left( r \right) -2\,\xi\,\alpha\,r+ rW^{\prime}+W \left( r \right) =0.
\end{eqnarray}
\end{widetext}

From Eq. (\ref{3.4}), one can obtain the density as follows (see Appendix)
\begin{equation}
\rho ={\frac{l\left[ rW^{\prime}-W\left( r\right) \right] }{\kappa \,{r}^{3}\left( 1+3\,w\right) }},
\label{rho11}
\end{equation}
in which $l=\xi \alpha $. Solving Eq. (\ref{3.6}) with Eq. (\ref{rho11}), we
find the shape function as follows 
\begin{equation}
W(r)=\frac{lr}{l+1}+\frac{r_{0}}{l+1}\left( \frac{r_{0}}{r}\right) ^{\frac{5wl+3l+3w+1}{wl-l+3w+1}},  \label{wh11}
\end{equation}
where $W(r_{0})=r_{0}\neq 0:$ throat radius. Inserting Eq. (\ref{wh11}) into
Eq. (\ref{3.5}), we get a condition on $\omega $ as the following 
\begin{equation}
w=-\frac{l+1}{5\,l+3}.
\end{equation}Thus, the shape function and density become 
\begin{equation}
W(r)={\frac{1}{l+1}\left( lr+\mathit{r_{0}}\,\left( {\frac{\mathit{r_{0}}}{r}}\right) ^{{\frac{-5\,l-3}{3\,l+1}}}\right) },  \label{eqW}
\end{equation}\begin{equation}
\rho ={\frac{5l+3}{\kappa \left( 3l+1\right) {{\mathit{r_{0}}}}^{{{\mathit{2}}}}}}\left( r^{2}{{\mathit{r_{0}}}}\right) ^{{\frac{-2l}{3l+1}}}.
\label{eqW22}
\end{equation}It can be easily checked that as $l\rightarrow 0$ $\Longrightarrow \
w\rightarrow -\frac{1}{3},$ $W(r)\rightarrow \frac{r^{3}}{{{\mathit{r_{0}}}}^{{{\mathit{2}}}}},$ and $\rho \rightarrow \frac{3}{\kappa {{\mathit{r_{0}}}}^{{{\mathit{2}}}}}$. On the other hand, one can easily see that $g_{rr}$
diverges at$\ r=r_{0}$. Furthermore, Eq. (\ref{eqW}) is non-asymptotically
flat when ($r\rightarrow \infty $):
\begin{equation}
\lim_{r\rightarrow \infty }\frac{W(r)}{r}\rightarrow \frac{l}{l+1}+\lim_{r\rightarrow \infty } \left( \frac{r_0}{r}\right)^{1+\frac{-5l-3}{3l+1}}. \label{eq11}
\end{equation}
From the above equation, we infer that the first term is independent of $r$,
while the second term is vanished if $1+\frac {-5\,l-3}{3\,l+1}>0$. Non-asymptotically
flatness reflects the non-trivial topological structure (arising from the LSB
effects) of the wormhole. Such solutions are similar to the spacetimes whose having topological defects and dilaton
fields \cite{conical,nas1,nas2}. However, in principle, the solution
obtained should be matched to an exterior vacuum solution (for details, a
reader may refer to \cite{MontelongoGarcia:2010xd}).

The Ricci scalar results in 
\begin{equation}
R=\frac{3W^{\prime }{r}^{2}-2W\left( r\right) W^{\prime }+3W\left(r\right)
^{2}}{2{r}^{6}}.  \label{myiz}
\end{equation}

At $r=r_{0}$, Eq. (\ref{myiz}) results in:

\begin{equation}
R|_{r_{0}}={\frac{18{l}^{2}+24l+12}{{\mathit{r_{0}}}^{4}\left( 3\,l+1\right)
^{2}}}.  \label{yes1}
\end{equation}

In a similar way, the Kretschmann scalar: 
\begin{equation}
K=2\,{\frac{{r}^{2}W^{\prime 2}-2rW\left( r\right) W^{\prime }+3W\left(
r\right)^{2}}{{r}^{6}}},
\end{equation}%
yields 
\begin{equation}
K|_{r_{0}}={\frac{36\,{l}^{2}+24\,l+12}{{\mathit{r_{0}}}^{4}\left(
3\,l+1\right) ^{2}}}.  \label{yes2}
\end{equation}%
It is clear from Eqs. (\ref{yes1}) and (\ref{yes2}) that singularity arises
if $l=-\frac{1}{3}$.

\subsection{Flare-out conditions:}

Traversability of a wormhole is determined by the flare-out conditions \cite%
{Travel}. We can easily visualize the spatial geometry of the wormhole using
an embedding diagram. The metric with $t=t_{0}$ (constant) reduces to the
following form at the equatorial plane $\theta =\frac{\pi }{2}:$ \cite%
{Muller:2008zza} 
\begin{equation}
ds^{2}=\frac{dr^{2}}{1-\frac{W(r)}{r}}+r^{2}d\varphi ^{2}.
\end{equation}

Then, we embed the wormhole geometry into a Euclidean 3-space:

\begin{equation}
d\sigma ^{2}=dz^{2}+dr^{2}+r^{2}d\varphi ^{2},
\end{equation}%
which can be rewritten as follows 
\begin{equation}
d\sigma^{2}=(1+{z^{\prime}}^2)dr^{2}+r^{2}d\varphi ^{2},
\end{equation}%
where 
\begin{equation}
z^{\prime}=\pm \frac{1}{\sqrt{\frac{r}{W(r)}-1}}.
\end{equation}%
We can now calculate the proper radial distance, which ought to be real and
finite: 
\begin{equation}
d(r)=\int_{r_0}^{r}\frac{dr}{\sqrt{1-\frac{W(r)}{r}}}.
\end{equation}

We deduce from the above equation that 
\begin{equation}
\sqrt{1-\frac{W(r)}{r}}>0.
\end{equation}

It is worth noting that there is a coordinate singularity at the throat of
the wormhole. Thus, the flare-out conditions \cite{Travel,Zangeneh:2015jda}
yield:

\begin{equation}
W(r)-r \leq 0,
\end{equation}

and 
\begin{equation}
rW^{\prime}-W(r)<0.
\end{equation}%
Thus, we have 
\begin{equation}
W^{\prime }={\frac {3\,l+3}{3\,l+1}}<1. \label{myflr}
\end{equation}

\begin{figure}[h!]
\center
\includegraphics[width=0.4\textwidth]{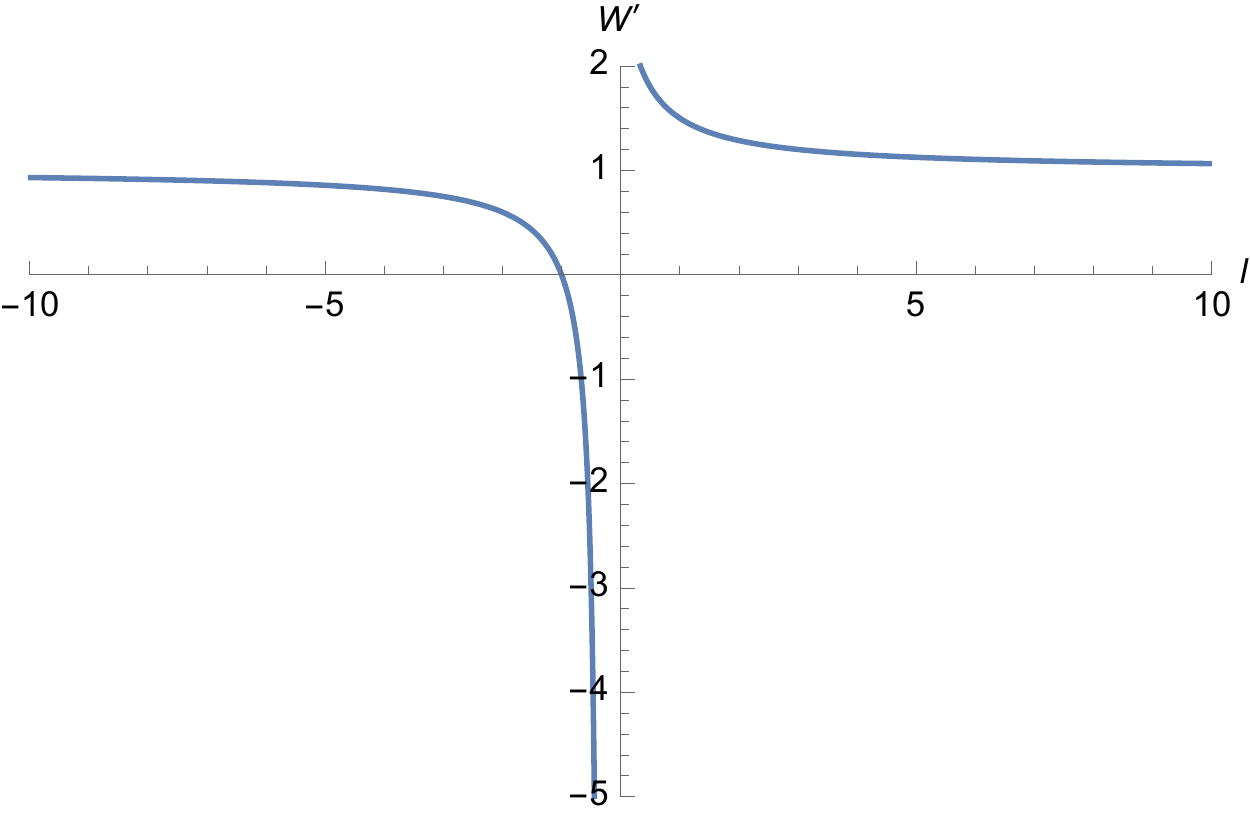}
\caption{ $W^{\prime}$ versus $l$ graph. The flare-out conditions are satisfied for the negative values of l.}
\label{FL0}
\end{figure}

For the condition (\ref{myflr}), it is depicted in Fig. (\ref{FL0}) that the flare-out conditions are
satisfied for the negative $l$ values.

\section{Energy conditions}

Following the monograph of \cite{Wald:1984rg}, in this section, we shall analyze the energy conditions for the bumblebee wormhole described with Eqs.(\ref{mymetric}) and
(\ref{eqW}). \newline

\subsection{Null energy condition:}

The null energy condition is expressed in terms of energy density and pressure as follows
\begin{equation}
\rho + p \geq 0,
\end{equation}

which yields

\begin{equation}
\rho + p ={\frac { \left( 4\,l+2 \right) \mathit{r_0}}{ \left( 3\,l+1 \right) \kappa \,%
{r}^{3}} \left( {\frac {\mathit{r_0}}{r}} \right) ^{{\frac {-5\,l-3}{3\, l+1}%
}}}\geq 0. \label{mynull}
\end{equation}
At $l=-\frac{1}{2}$, the null energy condition (\ref{mynull}) becomes zero. 
\begin{figure}[h]
\center
\includegraphics[width=0.5\textwidth]{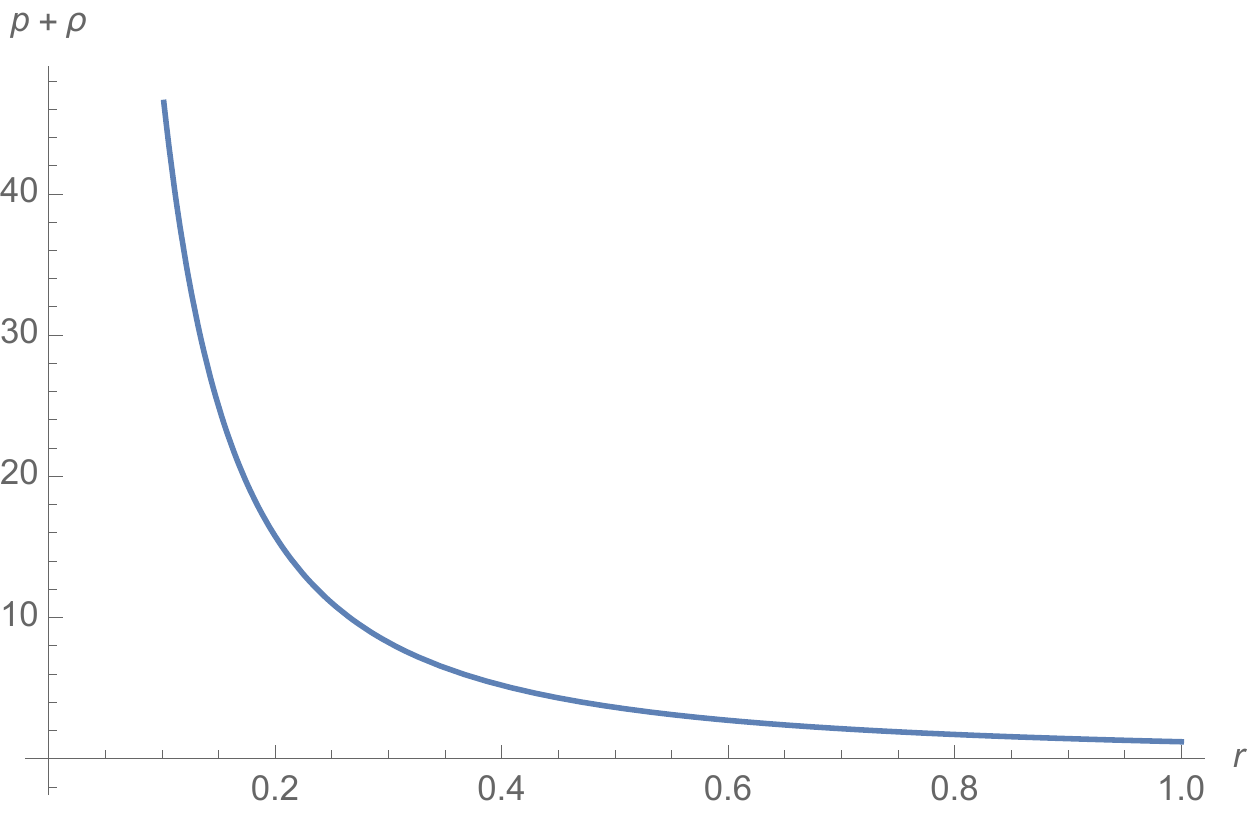}
\caption{Null energy condition $\protect\rho + P \geq 0$ is satisfied for $%
r_0=1$, $l=-2$ and $\protect\kappa=1$.}
\label{rhop1}
\end{figure}
It is clear from Fig. (\ref{rhop1}) that the null energy condition for the
bumblebee wormhole is satisfied. Moreover, we depict an interactive plot in 
\cite{intrhop} for the null energy condition of the bumblebee wormhole in
order to present the effect of parameter $l$.

\begin{figure}[h!]
\center
\includegraphics[width=0.5\textwidth]{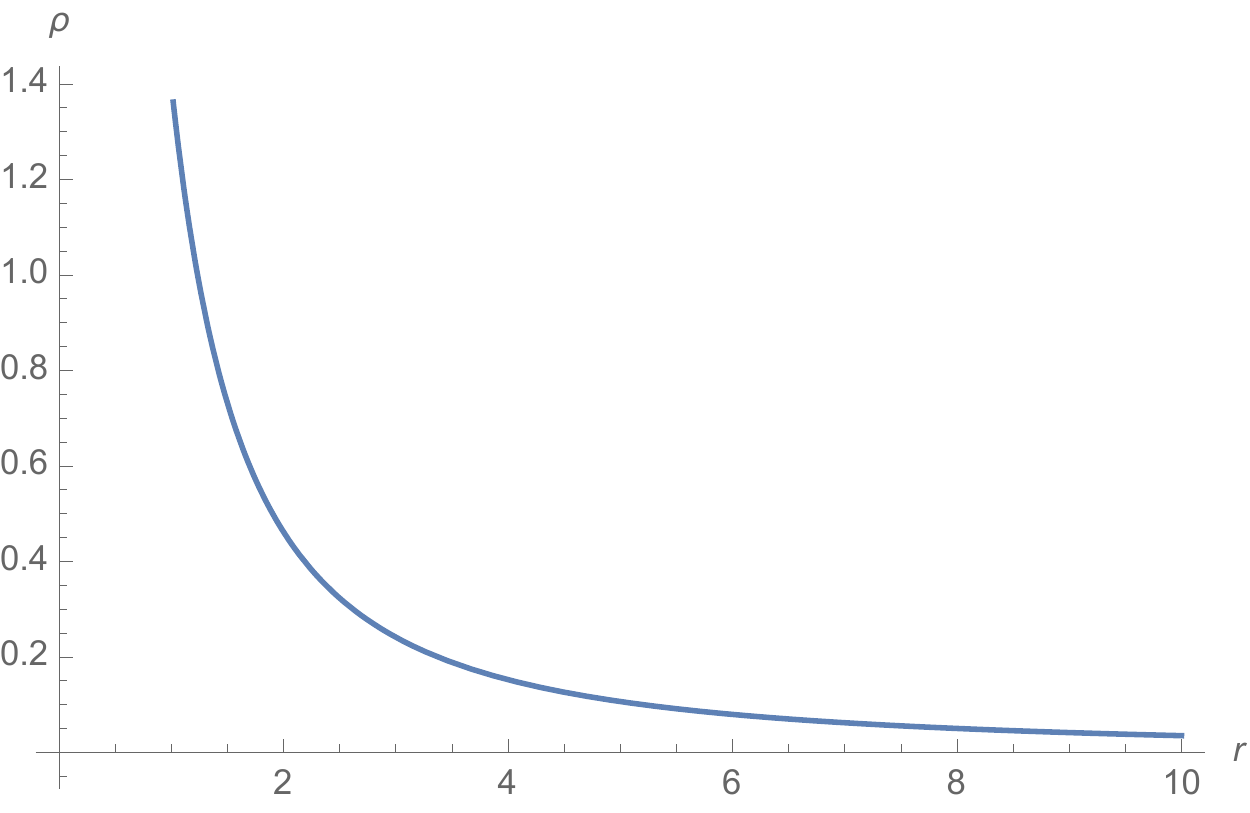}
\caption{The plot of energy density $\protect\rho$ for the values of
parameters $r_0=1$, $l=-2$, and $\protect\kappa=1$.}
\label{rho1}
\end{figure}

\subsection{Weak energy condition:}

Weak energy condition is given by 
\begin{equation}
\rho \geq 0\,, \quad \rho + p \geq 0,
\end{equation}

which gives the following result for the bumblebee wormhole:

\begin{equation}
\rho={\frac {\mathit{r_0}\, \left( 5\,l+3 \right) }{ \left( 3\,l+1 \right)
\kappa\,{r}^{3}} \left( {\frac {\mathit{r_0}}{r}} \right) ^{{\frac {-5\,l- 3%
}{3\,l+1}}}} \geq 0.
\end{equation}

In Fig. (\ref{rho1}), we show that weak energy condition for the
bumblebee wormhole is satisfied when the physical parameters are fixed to $%
r_0=1$, $l=-2$, and $\kappa=1$. One can reach to the interactive plot of the weak energy condition for the bumblebee wormhole with the link given in \cite{intrhop}. By
this way, the effect of $l$ on the weak energy condition can be monitored.

\subsection{Strong energy condition:}

Strong energy condition is governed by 
\begin{equation}
\rho + 3p \geq 0\,, \quad \rho + p \geq 0,
\end{equation}

which yields the following expression for the wormhole of bumblebee gravity:
\begin{equation}
\rho + 3p={\frac {{2 \mathit{r_0}}\,l}{ \left( 3\,l+1 \right) \kappa\,{r}%
^{3}} \left( {\frac {\mathit{r_0}}{r}} \right) ^{{\frac {-5\,l-3}{3\,l+1}}}}
\geq 0
\end{equation}

\begin{figure}[h!]
\center
\includegraphics[width=0.5\textwidth]{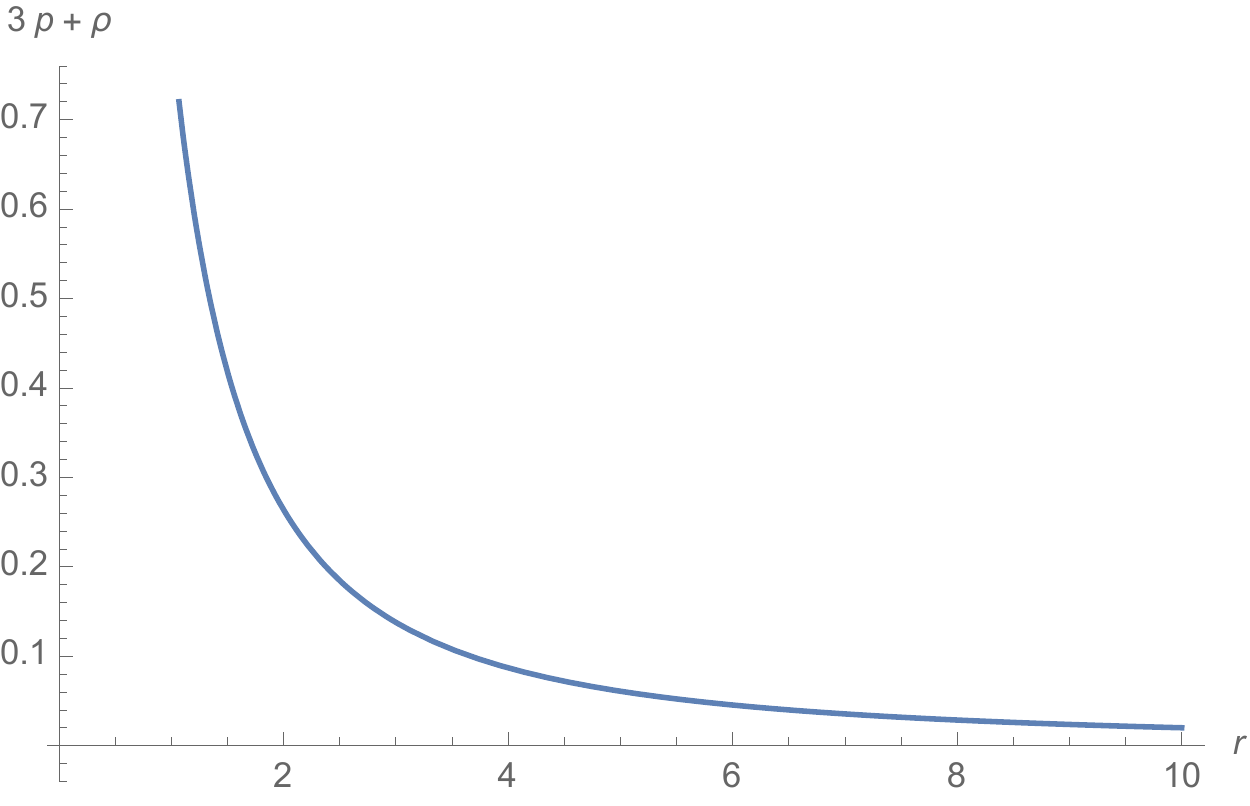}
\caption{The plot of $\protect\rho+3P$ is satisfied for $r_0=1$, $l=-2$ and $%
\protect\kappa=1$.}
\label{rho3p}
\end{figure}

From Fig. (\ref{rho3p}), it can be seen that strong
energy condition for the bumblebee wormhole is satisfied for the parameters
of $r_0=1$, $l=-2$, and $\kappa=1$. To reach to the interactive plot of the strong energy
condition for the bumblebee wormhole, one can follow the link given in \cite{intrho3p}.

\section{Deflection of light}

In this section, we shall explore the effect of bumblebee gravity in the
gravitational lensing of the spacetime of the wormhole metric described by
Eqs. (\ref{mymetric}) and (\ref{wh11}). For simplicity, the Lagrangian is
chosen in the equatorial plane. Thus, we get 
\begin{equation}
2\,\mathcal{L}=-\dot{t}^{2}+\frac{(1+l)\dot{r}^{2}}{1-\left( \frac{r_{0}}{r}%
\right) ^{1+\frac{5wl+3l+3w+1}{wl-l+3w+1}}}+r^{2}\dot{\varphi}^{2}.
\end{equation}

There are two constants of motion (energy and angular momentum) for a
massless particle, which are defined as follows 
\begin{eqnarray}
E &=&-g_{\mu \nu }K^{\mu }U^{\nu }=\frac{dt}{d\lambda }, \\
L &=& g_{\mu \nu }\Phi ^{\mu }U^{\nu }=r^{2}\frac{d\varphi }{d\lambda },
\end{eqnarray}%
in which $\lambda $ denotes the affine parameter along the light ray. Note
that $K^{\mu }$ and $\Phi ^{\mu }$ are the timelike and rotational Killing
vectors, respectively. One can define the impact parameter of the light ray
as 
\begin{equation}
b=\frac{L}{E}=r^{2}\frac{d\varphi }{dt}.
\end{equation}

From the above relations, one can find the following differential equation
for the light ray 
\begin{equation}
\left( \frac{dr}{d\varphi }\right) ^{2}+\frac{r^{2}}{B(r)}=\frac{r^{4}}{%
b^{2}B(r)},
\end{equation}%
in which 
\begin{equation}
B(r)=\frac{(1+l)}{1-\left( \frac{r_{0}}{r}\right) ^{1+\frac{5wl+3l+3w+1}{%
wl-l+3w+1}}}.
\end{equation}

One can solve this equation by introducing a new variable, let us say $%
u(\varphi )$, which is related with the radial coordinate as $r=\frac{1}{%
u(\varphi )}$. If we use the following identity: 
\begin{equation}
\frac{\dot{r}}{\dot{\varphi}}=\frac{\mathrm{d}r}{\mathrm{d}\varphi }=-\frac{1%
}{u^{2}}\frac{\mathrm{d}u}{\mathrm{d}\varphi },  \label{iden1}
\end{equation}%
then in the large $r$ limit, it is possible to recover the following
relation 
\begin{equation}
\frac{d^{2}u}{d\varphi ^{2}}+\beta u=0.  \label{iden2}
\end{equation}

Furthermore, $\beta =(1+l)^{-1}$, since in the weak limit we have the
following approximation: $B(r)\rightarrow 1+l$ as $r\rightarrow \infty $.
The solution of the last differential equation is given by 
\begin{equation}
u(\varphi )=A_{1}\sin (\sqrt{\beta }\varphi )+A_{2}\cos (\sqrt{\beta }%
\varphi ).  \label{iden3}
\end{equation}

When using the following initial conditions $u(\varphi =0)=0$ and $u(\varphi
=\pi /2)=\frac{1}{b}$, we find 
\begin{equation}
u(\varphi )=\frac{\sin (\sqrt{\beta }\varphi )}{b}\left( \sin (\frac{\sqrt{%
\beta }\pi }{2})\right) ^{-1}.  \label{iden4}
\end{equation}

Moreover, one can use $\sin (\frac{\sqrt{\beta }\pi }{2})\simeq 1$ and in
sequel derives the light ray expression: 
\begin{equation}
r=\frac{b}{\sin (\sqrt{\beta }\varphi )}.  \label{iden5}
\end{equation}

This equation is important in computing the deflection angle in the GBT.
Next, let us find the wormhole optical metric by letting $\mathrm{d}s^{2}=0$%
, which corresponds to 
\begin{equation}
\mathrm{d}t^{2}=\frac{(1+l)\,dr^{2}}{1-\left( \frac{r_{0}}{r}\right) ^{1+%
\frac{5wl+3l+3w+1}{wl-l+3w+1}}}+r^{2}\mathrm{d}\varphi ^{2}.
\end{equation}

It is also possible to write down the wormhole optical metric in terms of
new coordinates $x^{a}$: 
\begin{equation}
\mathrm{d}t^{2}=h_{ab}\,\mathrm{d}x^{a}\mathrm{d}x^{b}=\mathrm{d}\zeta ^{2}+%
\mathcal{H}^{2}(\zeta )\mathrm{d}\varphi ^{2},
\end{equation}%
where 
\begin{equation}
\mathrm{d}\zeta =\frac{\sqrt{1+l}\,\mathrm{d}r}{\sqrt{1-\left( \frac{r_{0}}{r%
}\right) ^{1+\frac{5wl+3l+3w+1}{wl-l+3w+1}}}},\,\,\,\mathcal{H}=r.
\end{equation}

The Gaussian optical curvature (GOC) $\mathcal{K}$ can be found to be (see
for details, \cite{gibbons}) 
\begin{eqnarray}
\mathcal{K} &=&-\frac{1}{\mathcal{H}}\left[ \frac{\mathrm{d}r}{\mathrm{d}%
\zeta }\frac{\mathrm{d}}{\mathrm{d}r}\left( \frac{\mathrm{d}r}{\mathrm{d}%
\zeta }\right) \frac{\mathrm{d}\mathcal{H}}{\mathrm{d}r}+\left( \frac{%
\mathrm{d}r}{\mathrm{d}\zeta }\right) ^{2}\frac{\mathrm{d}^{2}\mathcal{H}}{%
\mathrm{d}r^{2}}\right]  \notag \\
&=&-\frac{\left( 1+\Xi \right) }{2\,r^{2}\,\left( 1+l\right) }\left( \frac{%
r_{0}}{r}\right) ^{1+\Xi },
\end{eqnarray}%
where 
\begin{equation}
\Xi =\frac{5wl+3l+3w+1}{wl-l+3w+1}.
\end{equation}%
Alternatively, one can approximate the above equation by expanding in series
around $l$. Thus, we get 
\begin{equation}
\mathcal{K}\simeq -\frac{r_{0}^{2}}{r^{4}}-\frac{r_{0}^{2}\,l\left[ 4\ln (%
\frac{r_{0}}{r})(w+1)-w-1\right] }{r^{4}(3w+1)}.
\end{equation}

The key point in this method is that a non-singular domain outside the light
ray, say $\mathcal{A}_{R}$, which is bounded by $\partial \mathcal{A}%
_{R}=\gamma _{h}\cup C_{R}$ should be chosen. The GBT in the context of the
optical geometry is expressed as follows 
\begin{equation}
\iint\limits_{\mathcal{A}_{R}}\mathcal{K}\,\mathrm{d}\sigma
+\oint\limits_{\partial \mathcal{A}_{R}}\kappa \,\mathrm{d}t+\sum_{k}\psi
_{k}=2\pi \chi (\mathcal{A}_{R}),  \label{10}
\end{equation}

in which $\kappa $ gives the geodesic curvature, $\mathrm{d}\sigma $ is the
optical surface element, and $\psi _{k}$ stands for the exterior angle at
the $k^{th} $ vertex. We set the Euler characteristic number to one, i.e., $%
\chi (\mathcal{A}_{R})=1.$ Thus, the geodesic curvature is defined by \cite%
{gibbons}

\begin{equation}
\kappa =h\,\left( \nabla _{\dot{\gamma}}\dot{\gamma},\ddot{\gamma}\right) ,
\end{equation}

where the unit speed condition is selected as $h(\dot{\gamma},\dot{\gamma})=1
$. For a very large radial coordinate $R\rightarrow \infty $, our two jump
angles (at the source $\mathcal{S}$, and observer $\mathcal{O})$, yield $%
\psi _{\mathit{O}}+\psi _{\mathit{S}}\rightarrow \pi $ \cite{gibbons}. Thus,
the GBT simplifies to 
\begin{equation}
\iint\limits_{\mathcal{A}_{R}}\mathcal{K}\,\mathrm{d}\sigma
+\oint\limits_{C_{R}}\kappa \,\mathrm{d}t\overset{{R\rightarrow \infty }}{=}%
\iint\limits_{\mathcal{A}_{\infty }}\mathcal{K}\,\mathrm{d}\sigma
+\int\limits_{0}^{\pi +\hat{\alpha}}\mathrm{d}\varphi =\pi .  \label{12}
\end{equation}

By definition, the geodesic curvature for $\gamma _{h}$ is set to zero.
Then, we are left with a contribution from the curve $C_{R}$, which is
located at a distance $R$ from the wormhole center in the equatorial plane.
In short, we need to compute the following: 
\begin{equation}
\kappa (C_{R})=|\nabla _{\dot{C}_{R}}\dot{C}_{R}|.
\end{equation}

In component notation, the radial part can be written as 
\begin{equation}
\left( \nabla _{\dot{C}_{R}}\dot{C}_{R}\right) ^{r}=\dot{C}_{R}^{\varphi
}\,\left( \partial _{\varphi }\dot{C}_{R}^{r}\right) +\Gamma _{\varphi
\varphi }^{r(op)}\left( \dot{C}_{R}^{\varphi }\right) ^{2}.
\end{equation}

With the help of the unit speed condition, one can calculate the Christoffel
symbols that are related to our optical metric in the large coordinate
radius $R$ and gets

\begin{eqnarray}
\lim_{R\rightarrow \infty }\kappa (C_{R}) &=&\lim_{R\rightarrow \infty
}\left\vert \nabla _{\dot{C}_{R}}\dot{C}_{R}\right\vert ,  \notag \\
&\rightarrow &\frac{1}{\sqrt{1+l}\,R}.
\end{eqnarray}

To understand the meaning of the above equation, we rewrite the optical
metric for a constant $R$. Thus, we have

\begin{equation}
\lim_{R\rightarrow \infty }\mathrm{d}t\rightarrow R\,\mathrm{d}\varphi .
\end{equation}

Combining the last two equations together, we obtain 
\begin{equation}
\kappa (C_{R})\mathrm{d}t=\frac{1}{\sqrt{1+l}}\mathrm{d}\varphi .
\end{equation}

This equation implies that our wormhole geometry is non-asymptotically flat
and correspondingly, the optical metric is not asymptotically Euclidean.
Using this result, we can express the deflection angle as follows 
\begin{equation}
\hat{\alpha}=\left( \sqrt{1+l}-1\right) \pi -\sqrt{1+l}\int\limits_{0}^{\pi
}\int\limits_{\frac{b}{\sin \left( \frac{\varphi }{\sqrt{1+l}}\right) }%
}^{\infty }\mathcal{K}\mathrm{d}\sigma ,  \label{def}
\end{equation}%
where the light ray $r(\varphi )=\frac{b}{\sin \left( \frac{\varphi }{\sqrt{%
1+l}}\right) }$ ($b$ is now interpreted as the \textit{impact parameter \cite%
{kimet3}}) can be approximated to the closest distance that is obtained from
the wormhole in the first order approximation. The first term of Eq. (\ref%
{def}) can be approximated as 
\begin{equation}
\left( \sqrt{1+l}-1\right) \pi =\frac{l\pi }{2}-\frac{l^{2}\pi }{8}+...
\end{equation}

The surface can also be approximated to 
\begin{equation}
\mathrm{d}\sigma =\sqrt{h}\,\mathrm{d}\zeta \,\mathrm{d}\varphi \simeq \sqrt{%
1+l}\,r\,\mathrm{d}r\,\mathrm{d}\varphi .
\end{equation}

Finally, the total deflection angle is found to be 
\begin{equation}
\hat{\alpha}\simeq \frac{l\pi }{2}-\int\limits_{0}^{\pi }\int\limits_{\frac{b%
}{\sin \left( \frac{\varphi }{\sqrt{1+l}}\right) }}^{\infty }\left[ -\frac{%
\left( 1+\Xi \right) }{2r}\left( \frac{r_{0}}{r}\right) ^{1+\Xi }\right] 
\mathrm{d}r\mathrm{d}\varphi .
\end{equation}

Evaluating the last integral, we find

\begin{equation}
\hat{\alpha}\simeq \frac{l\pi }{2}+\frac{\sqrt{\left( 1+l\right) \pi }}{2}%
\left( \frac{r_{0}}{b}\right) ^{1+\Xi }\frac{\Gamma \left( \frac{2+\Xi }{2}%
\right) }{\Gamma \left( \frac{3+\Xi }{2}\right) },
\end{equation}%
with the condition of $1+\Xi >0$. Performing a series expansion, we can
write the last equation as follows 
\begin{eqnarray}
\hat{\alpha} &\simeq &\frac{l\pi }{2}+\frac{\pi r_{0}^{2}}{4\,b^{2}}+\frac{%
5\pi r_{0}^{2}l}{8b^{2}}-\frac{\pi r_{0}^{2}lw}{b^{2}}+\frac{\pi
r_{0}^{2}l\ln (\frac{r_{0}}{b})}{b^{2}}  \notag \\
&-&\frac{\pi r_{0}^{2}l\ln 2}{b^{2}}+\frac{2\pi r_{0}^{2}lw\ln 2}{b^{2}}-%
\frac{2\pi r_{0}^{2}lw\ln (\frac{r_{0}}{b})}{b^{2}}.
\end{eqnarray}

Note that for vanishing bumblebee gravity, $l=0$, it reduces to the original
Einstein's gravity and whence the deflection angle becomes $\hat{\alpha}%
\simeq \,{\frac{\pi \,{\mathit{r_{0}}}^{2}}{4{b}^{2}}}${,} which is in
agreement with the Ellis wormhole \cite{wh10}. On the other hand, if $1+\Xi
\leq 0$, we can only incorporate the finite distance corrections in the
deflection angle of light.

It is worth to re-emphasize that due to the LSB effect, there are additional terms seen in the left hand side of Eq. (\ref{eq11}) that yields a non-asymptotically flat spacetime. Although the first term of Eq. (\ref{eq11}) is independent from the radial coordinate, however the second term should be vanished when $r\to \infty$, which is achieved by the condition of $1+\frac {-5\,l-3}{3\,l+1}>0$. Otherwise, the second term blows up, then the GBT becomes problematic. In fact, following paper of Ishihara et al \cite{asahi1}, one can only apply a finite correction to the deflection of light. Furthermore, there is a reference paper \cite{kimet3} in which a similar work was carried out for a different spacetime. To conclude, in a certain framework,  the GBT can be applied to the spacetimes in the presence of LSB effects.

\section{Conclusion}

We searched for a way to construct a traversable wormhole solution, one
which satisfies the energy conditions to become the most interesting
application of the general relativity theory. In finding some realistic
matter source that keeps the wormhole throat open such that interstellar or
inter-universe travel might become possible, the modified theories of
gravity are thought to be new remedy. We have, therefore, considered the
bumblebee gravity model to have such a traversable wormhole solution that
satisfies the null, weak, and strong energy conditions. In this paper, we
first derived the modified Einstein's field equations for the Lorentzian
wormhole in the bumblebee gravitational field. Next, using the associated
field equations with bumblebee gravity, we have obtained the new traversable
wormhole solution with the exact shape function (\ref{eqW}) and with $w=-%
\frac{l+1}{5\,l+3}$. Then, physical features of the obtained wormhole were
studied in detail. Singularity of the solution was analyzed by computing the
Ricci and Kretschmann scalars. It is seen that the singularity appears when $%
l=-\frac{1}{3}$. Afterwards, we have checked the flare-out conditions $%
W^{\prime }<1$ for the obtained bumblebee wormhole solution. We have shown
that the flare-out conditions are satisfied if $\frac {3\,l+3}{3\,l+1}<1$
where it is plotted in Fig. (\ref{FL0}).

In section IV, we checked the energy conditions (null, weak, and strong) for
the bumblebee wormhole and rendered them graphically. In Figs. (\ref{rhop1},%
\ref{rho1},\ref{rho3p}), we analyzed the three energy conditions of the
bumblebee wormhole for the values of $r_0=1$, $l=-2$ and $\kappa=1$. We
noted that all energy conditions for the bumblebee wormhole were satisfied
when $r_0=1$, $l=-2$ and $\kappa=1$. We also plotted the energy conditions
manipulated as interactive in \cite{intrhop,intrho,intrho3p}. 

Another important point is that under the LSB effect, the global topology of
the wormhole spacetime changes. The limit of $\frac{W(r)}{r}$ at spatial
infinity was found to be $\left. \frac{W(r)}{r}\right\vert _{r\rightarrow
\infty }\rightarrow \frac{l}{l+1}$, which shows that our wormhole solution
was non-asymptotically flat. The deflection of light was computed by
applying the GBT to the bumblebee wormhole expressed in the optical
geometry. It was seen that bumblebee parameter affects the geodesic optical
curvature, modifying the final result for the deflection angle. Due to the
nontrivial global topology, we have shown that the deflection of light is
changed by $\delta \hat{\alpha}=l\pi /2$, which is purely a topological term
and independent of the impact factor $b$. In addition, we incorporated the
bumblebee effects in the total deflection angle by deriving the light ray
equation, which modifies the straight line approximation in the domain of
integration. In other words, the total deflection angle not only depends on
the geometry of the bumblebee traversable wormhole (i.e., throat radius $%
r_{0}$), but it varies with the coupling constant $l$, and state parameter $w
$. Finally, in the case of $l=0$, we recovered the Ellis wormhole deflection
angle as being reported in the literature \cite{wh10}.

In short, the bumblebee wormhole that we constructed satisfies the energy
conditions for normal matter and flare-out conditions near the throat. In
the near future, we plan to add new sources (scalar, electromagnetic etc.)
to the bumblebee gravity. In this manner, we wish to obtain new spacetime
solutions and analyze their physical features \cite{bias}.

\acknowledgments 
We are thankful to the Editor and anonymous Referees for
their constructive suggestions and comments. A. \"{O}. is grateful to
Institute for Advanced Study, Princeton for hospitality. A. \"{O}. thanks to
Prof. Eduardo Guendelman for valuable discussions. This work was supported
by the Chilean FONDECYT Grant No. 3170035 (A. \"{O}.). We sincerely thank the Prof. Ted Jacobson
 for constructive criticisms and valuable comments. 

\appendix
\center{\textbf{APPENDIX}}
\\
The generic line-element of the static and a spherically symmetric traversable
wormhole can be expressed as follows%

\begin{equation}
ds^{2}=-dt^{2}+e^{2\nu\left(  r\right)  }dr^{2}+r^{2}\left(  d\theta^{2}%
+\sin^{2}\theta d\varphi^{2}\right)  ,\tag{A1}\label{A1}%
\end{equation}

which has following non-zero Ricci tensors:
\begin{equation}
R_{rr}=\frac{2\nu^{\prime}}{r}{,}\tag{A2}\label{A2}%
\end{equation}

\begin{equation}
R_{\theta\theta}=\frac{R_{\varphi\varphi}}{\sin^{2}\theta}=1+{\frac
{r\nu^{\prime}-1}{{\mathrm{e}^{2\nu\left(  r\right)  }}}.}\tag{A3}\label{A3}%
\end{equation}

Recall that the prime symbol denotes the derivative with respect to variable
$r$. Setting%

\begin{equation}
b_{\mu}=\left[  0,b(r),0,0\right]  ,\tag{A4}\label{A5}%
\end{equation}

and making straightforward calculations, one can obtain the following results
from Eq. (\ref{meins1}):

\begin{widetext}
\begin{equation}
E_{tt}^{instein}=\left(  3\omega+1\right)  \rho r\kappa\,{\mathrm{e}%
^{4\,\nu\left(  r\right)  }}-2b\left(  r\right)  b^{\prime}\xi-\psi(r)\xi
r,\tag{A5}\label{A6}%
\end{equation}

\begin{equation}
E_{rr}^{instein}=r\rho\kappa\left(  \omega-1\right)  {\mathrm{e}%
^{4\,\nu\left(  r\right)  }}+4\,{\mathrm{e}^{2\,\nu\left(  r\right)  }}%
\nu^{\prime}-6\,b\left(  r\right)  b^{\prime}+12\,b\left(  r\right)  ^{2}%
\nu^{\prime}\xi-3\psi(r)\xi, \tag{A6}  \label{A77} 
\end{equation}

\begin{align}
E_{\theta\theta}^{instein} & \equiv E_{\varphi\varphi}^{instein}=\left(
2+\rho\kappa\left(  \omega-1\right)  {r}^{2}\right)  {\mathrm{e}%
^{4\,\nu\left(  r\right)  }}+2\left(  r\nu^{\prime}-1\right)  {\mathrm{e}%
^{2\,\nu\left(  r\right)  }}\nonumber\\
& -2\left(  -3\,b\left(  r\right)  ^{2}\nu^{\prime}r+3\,b\left(  r\right)
b^{\prime}r+b \left(  r\right)^{2} \right)  \xi+r\psi(r)\xi,\tag{A7}%
\label{A8N}%
\end{align}

where

\begin{equation}
\psi(r)=\left[  3\,b\left(  r\right)  ^{2}\left(  \nu^{\prime}\right)
^{2}-b\left(  r\right)  ^{2}\nu^{\prime\prime}-5\,b\left(  r\right)
\nu^{\prime}b^{\prime}+b\left(  r\right)  b^{\prime\prime}+\left(  b^{\prime
}\right)  ^{2}\right]  r.\tag{A8}\label{A4}%
\end{equation}
\end{widetext}
One can immediately check that for the traversable wormhole metric (\ref{mymetric}) with the bumblebee vector (\ref{mymbumb}):%

\begin{equation}
\nu\left(  r\right)  =-\frac{1}{2}\ln\left(  1-\frac{W(r)}{r}\right)
,\tag{A9}\label{A9}%
\end{equation}

\begin{equation}
b\left(  r\right)  =\sqrt{\frac{\alpha}{1-\frac{W(r)}{r}}},\tag{A10}%
\label{A10}%
\end{equation}

Eq. (\ref{A4}) yields $\psi(r)=0$. For this reason, in the obtained Einstein field equations (\ref{3.4})-(\ref{3.6}), there are not any $W^{\prime2}$ and $W^{\prime\prime}$ terms.

\end{document}